\documentclass{elsart}
\usepackage{epsfig}

\begin{document}

\begin{frontmatter}

\title{Improvement of the extended P+QQ interaction
       by modifying the monopole field}

\author[Sawara]{M. Hasegawa}, \author[Higashi]{K. Kaneko} and
\author[Jonan]{S. Tazaki}
\address[Sawara]{Laboratory of Physics, Fukuoka Dental College,
 Fukuoka 814-0193, Japan}
\address[Higashi]{Department of Physics, Kyushu Sangyo University,
 Fukuoka 813-8503, Japan}
\address[Jonan]{Department of Applied Physics, Fukuoka University,
 Fukuoka 814-0180, Japan}

\begin{abstract}

  The extended pairing plus $QQ$ interaction with the $J$-independent
 isoscalar proton-neutron force as an average monopole field, which has
 succeeded in describing collective yrast states of $N \approx Z$
 even-$A$ nuclei, is improved.
 The improvement is accomplished by adding small monopole terms
 (relevant to spectroscopy) to the average monopole field
 (indispensable to the binding energy).  This modification extends
 the applicability of the interaction to nuclei with $N \approx 28$
 such as $^{48}$Ca, and moreover improves energy levels of
 noncollective states.
 The modified interaction successfully describes not only even-$A$
 but also odd-$A$ nuclei in the $f_{7/2}$ shell region.
  Results of exact shell model calculations in the model space
 ($f_{7/2}$, $p_{3/2}$, $p_{1/2}$) (its usefulness has been
 demonstrated previously) are shown for $A$=47, 48, 49, 50
 and 51 nuclei.
 
\vspace*{3mm}
\leftline{PACS:  21.60.-n;21.10.Dr;21.10.Hw;21.10.Re}
\end{abstract}

\end{frontmatter}

%\newpage
%=================================================================
\section{Introduction}

  An extended $P+QQ$ force model \cite{Hasegawa,Kaneko} has been
 applied to even-$A$ nuclei in the $f_{7/2}$ shell region, in the
 previous paper \cite{Hasegawa2} (hereafter referred to as (I)).  
 The exact shell model calculations have shown that the
 $P_0+P_2+QQ+V^0_{\pi \nu}$ interaction well reproduces the observed
 energy levels and $E2$ transition probabilities of the collective
 yrast states as well as the binding energies in the $A$=46, 48 and
 50 nuclei with $N \approx Z$.  The correspondence between theory and
 experiment is almost comparable to that attained by the full $fp$
 shell model calculations with realistic effective interactions
 \cite{Caurier,Caurier2,Martinez,Zamick,Svensson,Friessner,Lenzi}.
 
  The $P_0+QQ$ or $P_0+P_2+QQ$ force
 \cite{Kisslinger,Baranger,Kishimoto,Hara} has been regarded as a
 schematic interaction which represents typical correlations in nuclei.
 The results in (I), however, told us that the extended $P+QQ$ force
 accompanied by the $J$-independent proton-neutron ($p$-$n$) force
 $V^0_{\pi \nu}$ is more than a mere schematic interaction
 and reflects the important aspects of the real interaction.
  The interaction serves for quantitative description of
 nuclear structure in the $fp$ shell region.
 This conclusion is consistent with the discussion of Dufour and
 Zuker \cite{Dufour} that the residual part of a realistic effective
 interaction obtained after extracting the monopole terms is dominated
 by the multipole forces ($P_0$, $QQ$ {\it etc.}). 
 In our model, the $V^0_{\pi \nu}$ force nicely plays the role of
 the average monopole field and recovers the deficiency of
 the binding energy given by the $P_0+P_2+QQ$ force.
 
   The calculated results in (I), however, revealed two flaws of the
 $P_0+P_2+QQ+V^0_{\pi \nu}$ interaction: (1) noncollective states
 except the collective yrast states have a tendency to go down;
 (2) binding energies and level schemes become worse as $N$ separates
 from $Z$, especially when $N$ is close to 28 as $^{48}$Ca.
 These flaws remind us of the divergence from experiment
 when the Kuo-Brown (KB) interaction \cite{Kuo} is applied to systems
 of many valence nucleons $^{48}$Ca, $^{49}$Ca etc \cite{McGrory}.
  The divergence was cured by modifying the monopole components of
 the $T=0$ and $T=1$ interaction matrix elements \cite{Pasquini,Poves}.
 The power of the modified interaction KB3 \cite{Caurier,Poves}
 was shown in the exact shell model calculations
 \cite{Caurier,Caurier2,Martinez,Martinez2,Poves2}.
 The modified aspect of another effective interaction FPD6
 \cite{Richter} refers also to the monopole terms \cite{Brown}.
 These indicate the importance of the monopole terms in spectroscopy
 \cite{Dufour,Pasquini,Poves,Brown,Duflo}.
 Our interaction has the most important term $V^0_{\pi \nu}$ as the
 average monopole field but does not include the monopole terms which
 affect spectroscopy.  It is a natural next concern to investigate
 whether the additional monopole terms can remove or not the flaws
 mentioned above.
 
   For this purpose, we carry out exact shell model calculations
 in the same model space ($f_{7/2},p_{3/2},p_{1/2}$) as in (I),
 where the model space was shown to be useful when the average monopole
 field ($V^0_{\pi \nu}$) independent of the extent of space is properly
 treated \cite{Hasegawa2}.  We introduce additional monopole terms
 so that their forms are as simple as possible and the number of
 new parameters is minimal.
 The modified interaction gives us satisfactory results for the
 $f_{7/2}$ shell nuclei. This paper presents only a short review for
 even-$A$ nuclei, because the conclusions in (I) for the collective
 yrast states of $N \approx Z$ even-$A$ nuclei are hardly changed.
 Our major attention is devoted to odd-$A$ nuclei, the structure of
 which is expected to be sensitive to the effective interaction
 employed.
 There are successful works by Mart\'{\i}nez-Pinedo \textit{et al}.
 \cite{Martinez2,Poves2} in which the $A$=47 and 49 nuclei are
 exhaustively studied by the full $fp$ shell model with the KB3
 interaction.  We examine the quality of our model by comparing
 our results with theirs as well as experimental ones. 
 We shall also show that our model describes $A$=51 nuclei very well.

  In Section 2, the additional monopole terms are introduced into our
 model and new parameters are determined for $^{48}$Ca and $^{47}$Ti.
 Section 3 makes a report about the effects of the modification
 on even-$A$ nuclei.  Section 4 examines the results of calculations
 in the cross conjugate nuclei ($^{47}$Ti, $^{49}$V)
 and ($^{47}$V, $^{49}$Cr). The results for $A$=51 are shown
 in Section 5.  Concluding remarks are given in Section 6.

%===============================================================
\section{modification of the monopole filed}

\begin{figure}[b]
\begin{center}
    \epsfig{file=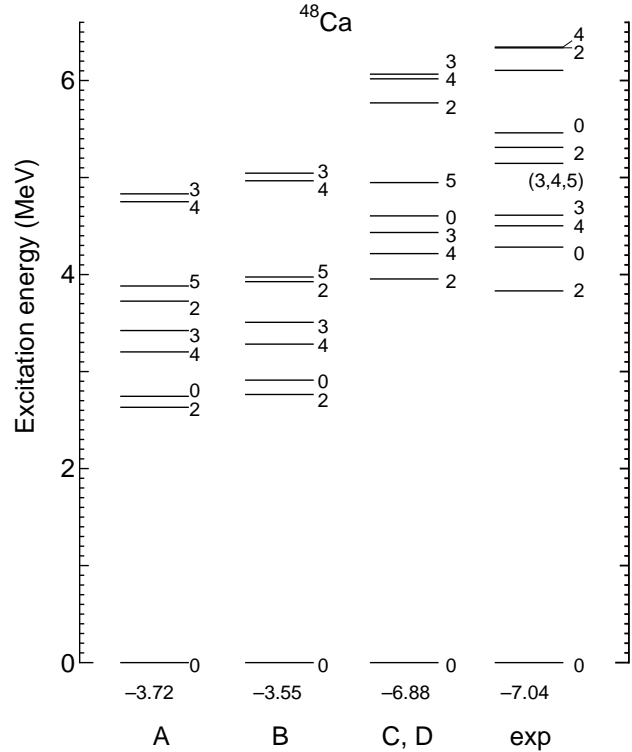,width=0.6\textwidth}
\caption{Energy levels of $^{48}$Ca calculated with four different sets
         of parameters, compared with the observed levels.
         The ground-state energy is also shown below the $0^+_1$ level
         in MeV.}
\label{fig1}
\end{center}
\end{figure}

  We start with the same model as in (I), {\it i.e.}, the
 $P_0+P_2+QQ+V^0_{\pi \nu}$ interaction in the model space
 $(f_{7/2},p_{3/2},p_{1/2})$ with the following parameter set
 called "Set A" in this paper (see (I) in detail):
\begin{eqnarray}
 \mbox{Set A: } &{}& \varepsilon_{7/2}=0.0,
        \quad \varepsilon_{3/2}=1.98,
        \quad \varepsilon_{1/2}=3.61, \nonumber \\
   &{}& g_0=0.48(42/A), \quad \ g_2=0.36(42/A)^{5/3}, \nonumber \\
   &{}& \chi^\prime=0.30(42/A)^{5/3}, \quad k^0=2.23(42/A)
         \quad \mbox{(in MeV).} \label{eq:1}
\end{eqnarray}
 This model succeeds in describing the collective yrast states of
 $N \approx Z$ nuclei, but cannot well reproduce binding energies and
 level schemes in nuclei with $N \approx 28$.  We consider $^{48}$Ca
 as a conspicuous example in Fig. \ref{fig1}.  The ground-state energy
 calculated with the parameter set A is -3.72 MeV against
 the experimental energy -7.04 MeV.  The obtained level scheme is shown
 in the first column A in Fig. \ref{fig1}. There is a significant
 discrepancy between the original model (column A) and experiment
 (column exp).

%===  Fig. 1 ========================================================
%====================================================================
 
  We used the naive single-particle energies with small spaces 
 from the observed levels of $^{41}$Ca in the Set A.  One may attribute
 the lowering of excited states to the small spaces
  $\epsilon_{7/2}-\epsilon_{3/2}$ and $\epsilon_{7/2}-\epsilon_{1/2}$.
  The single-particle energies, however, do not very much affect the
 level scheme in the $P_0+P_2+QQ$ force model, as mentioned in (I).
 If a little strengthened $QQ$ force is used for the single-particle
 energies of Kuo and Brown \cite{Kuo}, energy levels obtained are
 almost unchanged for the collective states in $N \approx Z$ nuclei. 
 The effect on $^{48}$Ca is shown in the second column B of Fig. 
 \ref{fig1}.  The small energy gap between $0^+_1$ (the ground state)
 and $2^+_1$ is little improved.
 The use of the Kuo-Brown single-particle energies which are often 
 employed will be convenient for comparison of our interaction with 
 other effective interactions.  We therefore use the new set of
 parameters (called "Set B") for the $P_0+P_2+QQ$ force in this paper,
\begin{eqnarray}
\mbox{Set B: } &{}& \varepsilon_{7/2}=0.0, \quad \varepsilon_{3/2}=2.1,
                     \quad \varepsilon_{1/2}=3.9, \nonumber \\
   &{}& g_0=0.48(42/A), \quad \ g_2=0.36(42/A)^{5/3}, \nonumber \\
   &{}& \chi^\prime=0.31(42/A)^{5/3}, \quad k^0=2.23(42/A)
         \quad \mbox{(in MeV).} \label{eq:2}
\end{eqnarray}

   In $^{48}$Ca, the dominant component of the ground state $0^+_1$ is
 the neutron configuration $(f_{7/2})^8$, and main components
 of the lowest excited states are $(f_{7/2})^7(p_{3/2})$ and
 $(f_{7/2})^7 (p_{1/2})$.  The deep binding energy of $0^+_1$ is
 probably due to the large correlation energy of the closed subshell
 configuration $(f_{7/2})^8$, which is given by
\begin{equation}
 \sum_{J=\mbox{even}}(2J+1) \langle f_{7/2}f_{7/2} J, T=1|
   V|f_{7/2}f_{7/2} J, T=1 \rangle .  \label{eq:3}
\end{equation}
 If we enlarge the value of Eq. (\ref{eq:3}), we can get a deep
 binding energy of $^{48}$Ca.  The enlargement of the $f_{7/2}$
 interaction matrix elements gives a smaller energy gain to the excited
 states than $0^+_1$, and hence extends the energy gap between $2^+_1$
 and $0^+_1$. The $2^+_1-0^+_1$ energy gap is further increased
 by weakening the interaction matrix elements
 $\langle f_{7/2} r J,T=1|V|f_{7/2} r J,T=1\rangle$
 ($r=p_{3/2}$ or $p_{1/2}$). 
  Note that only the isovector interactions with $T=1$ 
 take action in $^{48}$Ca.

  Therefore, a simple way to improve the binding energy and level 
 scheme of $^{48}$Ca is to add the following correction to the 
 isovector $(T=1)$ interactions:
\begin{equation}
  \sum_{ab} \sum_J \Delta k^1(abJ)
  \sum_{M\kappa} A^+_{JM1\kappa}(ab) A_{JM1\kappa}(ab), \label{eq:4}
\end{equation}
 where $A^\dagger_{JM\tau\kappa}(ab)=
 [c_a^\dagger c_b^\dagger]_{JM\tau\kappa}/\sqrt{1+\delta_{ab}}$.
 Neglecting the $J$-dependence of $\Delta k^1(abJ)$, in this 
 paper, we make further simplification 
\begin{equation}
 \Delta k^1(abJ)=
 \left\{
  \begin{array}{ll}
  \Delta k^1(f_{7/2}f_{7/2}) & \mbox{ for } a=b=f_{7/2} \\
  \Delta k^1(f_{7/2}r) & \mbox{ for } r=p_{3/2} \mbox{ or } p_{1/2} \\
  0  & \mbox{ for the others.} 
  \end{array}
 \right . \label{eq:5}                            
\end{equation}
Then, we have the isovector correction $\Delta V^1$ which belongs to
 the monopole terms discussed in Refs. \cite{Caurier,Dufour,Duflo},
\begin{eqnarray}
 \Delta V^1 &=&
  \Delta k^1(f_{7/2}f_{7/2}) \sum_{J=\mbox{even}} \sum_{M\kappa}
    A^\dagger_{JM1\kappa}(f_{7/2}f_{7/2}) A_{JM1\kappa}(f_{7/2}f_{7/2})
       \nonumber \\
&+&\Delta k^1(f_{7/2}r) \sum_{b \in r} \sum_{J} \sum_{M\kappa}
    A^\dagger_{JM1\kappa}(f_{7/2} b) A_{JM1\kappa}(f_{7/2} b).
       \label{eq:6}
\end{eqnarray}
We can expect $\Delta k^1(f_{7/2}f_{7/2})$ being attractive and
 $\Delta k^1(f_{7/2}r)$ being repulsive, from the above consideration.
 Trial calculations in $^{48}$Ca recommend the following corrections
 to the parameter set B, which we call "Set C":
\begin{eqnarray}
  \mbox{Set C: } &{}& \mbox{Set B accompanied by the corrections}
                      \nonumber \\
    &{}& \Delta k^1(f_{7/2}f_{7/2})=-0.14, \quad
    \Delta k^1(f_{7/2}r) = 0.05 \quad \mbox{(in MeV)}. \label{eq:7}
\end{eqnarray}
 The parameter set C yields the level scheme in the third column C
 of Fig. \ref{fig1}.  The agreement with the observed levels is
 satisfactory.  Figure \ref{fig1} testifies that the $P_0+P_2+QQ$ force
 is very much improved by adding the isovector monopole terms
 as Eq. (\ref{eq:6}).

%===  Fig. 2 ========================================================
\begin{figure}[b]
\begin{center}
    \epsfig{file=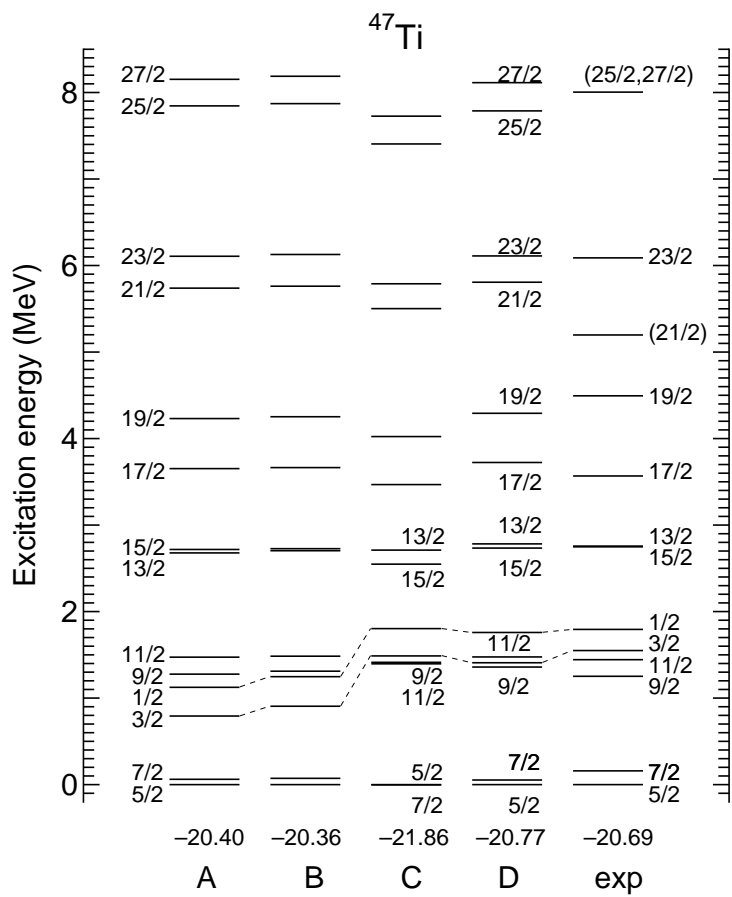,width=0.7\textwidth}
\caption{Energy levels of $^{47}$Ti calculated with four different sets
         of parameters, compared with the observed levels.
         The ground-state energy is also shown below the $0^+_1$ level
         in MeV.}
\label{fig2}
\end{center}
\end{figure}
%====================================================================

  Let us test the validity of $\Delta V^1$ with the parameters 
 (\ref{eq:7}) in odd-$A$ nuclei which are sensitive to the 
 interaction employed.  Figure \ref{fig2} illustrates the comparison
 between calculated yrast levels and observed ones in $^{47}$Ti,
 where only negative-parity states are considered in our model space
 $(f_{7/2}, p_{3/2}, p_{1/2})$.  The results in the columns A, B and
 C are obtained by using the parameter sets A, B and C, respectively.
 The results A and B show that the $P_0+P_2+QQ$ force fairly
 well reproduces the collective ground-state band on the $5/2^-$
 state but lays the noncollective low-spin states $3/2^-$ and 
 $1/2^-$ too much lower.  This tendency is one of the flaws of 
 the $P_0+P_2+QQ$ force.  The column C indicates that the isovector
 monopole terms in $\Delta V^1$ push the $3/2^-$ and $1/2^-$ states
 up to the correct positions.
  However, $\Delta V^1$ disturbs the order of the adjacent levels 
 $(5/2^-, 7/2^-)$ and $(9/2^-,11/2^-)$ at low energy
 and lowers the energies of high-spin states above $13/2^-$.
 Furthermore, the introduction of $\Delta V^1$ changes the binding
 energy of not only $^{48}$Ca but also the other nuclei. Especially,
 the additional isovector pairing interactions
 $\langle f_{7/2} f_{7/2} J,T=1|V|f_{7/2} f_{7/2} J, T=1 \rangle$ 
 operate on the main configurations of the ground states and cause 
 overbinding.

  These secondary troubles can be cured by introducing additional
 isoscalar $(T=0)$ monopole terms and by weakening the $J$-independent
 isoscalar force $V^0_{\pi \nu}$ (note that they are $p$-$n$
 interactions).  First, let us write the isoscalar correction
 $\Delta V^0$ in a form similar to $\Delta V^1$ 
\begin{eqnarray}
 \Delta V^0 &=&
  \Delta k^0(f_{7/2}f_{7/2}) \sum_{J=\mbox{odd}} \sum_M
    A^\dagger_{JM00}(f_{7/2}f_{7/2}) A_{JM00}(f_{7/2}f_{7/2})
     \nonumber \\
  &+& \Delta k^0 (f_{7/2}r) \sum_{b \in r} \sum_J \sum_M
    A^\dagger_{JM00}(f_{7/2}b) A_{JM00}(f_{7/2}b). \label{eq:8}
\end{eqnarray}
 We determine the parameters $\Delta k^0 (f_{7/2}f_{7/2})$ and
 $\Delta k^0(f_{7/2}r)$ so that the inverse order
 of $(5/2^-, 7/2^-)$ and $(9/2^-,11/2^-)$ are restored and
 the high-spin levels get near to the observed ones. 
 Secondly, we weaken the strength $k^0$ of $V^0_{\pi \nu}$ so as to
 obtain relatively good binding energies for nuclei from $A$=42 to
 $A$=51 as a whole. 
 The obtained parameter set, which we call "Set D", is
\begin{eqnarray}
  \mbox{Set D: } &{}& \mbox{Set C accompanied by the corrections}
                      \nonumber \\
    &{}& k^0 = 2.15(42/A), \nonumber \\
    &{}& \Delta k^0(f_{7/2}f_{7/2})=0.18, \quad
    \Delta k^0(f_{7/2}r) = -0.07 \quad \mbox{(in MeV)}. \label{eq:9}
\end{eqnarray}

%===  Fig. 3 ========================================================
\begin{figure}[b]
\begin{center}
    \epsfig{file=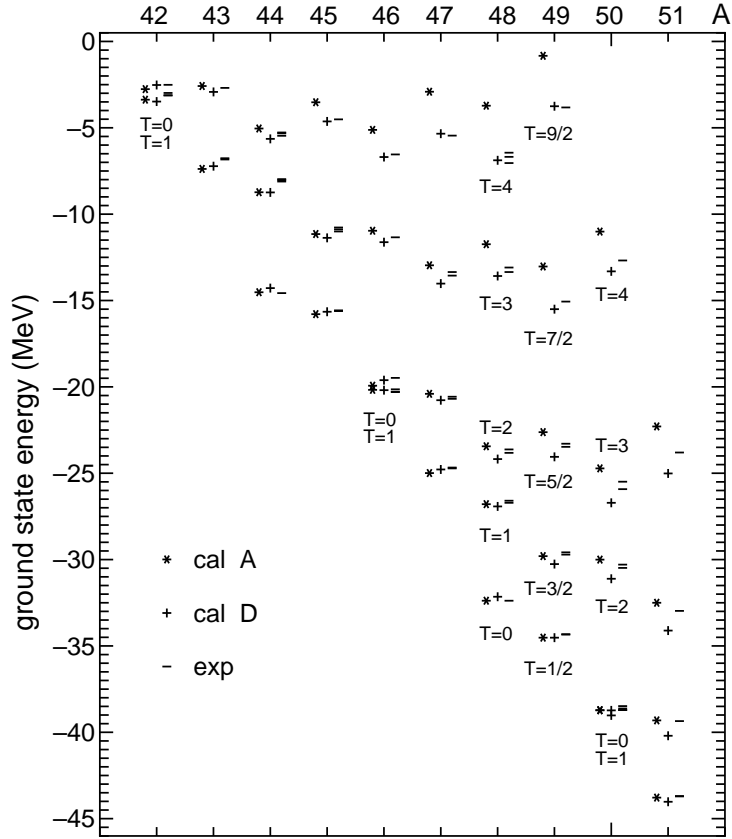,width=0.7\textwidth}
\caption{Ground-state energies of $A$=42 - 51 nuclei calculated with
         the original model A and modified one D, and the experimental
         energies. The experimental energies of the corresponding
         isobaric analogue states are also plotted.  For instance,
         the $T=4$ states at $A$=48 show the ground state of $^{48}$Ca
         and the $T$=4, 0$^+~$ states of $^{48}$Sc and $^{48}$Ti.}
\label{fig3}
\end{center}
\end{figure}
%====================================================================

  The parameter set D yields the level scheme in the fourth
 column D of Fig. \ref{fig2}.
 The energy levels of the ground-state band are well reproduced
 as those of the original model A and the $3/2^-$ and $1/2^-$ levels
 lie at good positions.  The agreement with the observed levels is
 considerably good.  The $p$-$n$ interactions $\Delta V^0$
 and $V^0_{\pi \nu}$ do not affect the spectrum of the $^{48}$Ca
 nucleus which has only valence neutrons.  The third column of
 Fig. \ref{fig1} for $^{48}$Ca is nothing but the result obtained
 by using the parameter set D.
  We can determine the strengths of $\Delta k^0 (f_{7/2}f_{7/2})$ and
 $\Delta k^0(f_{7/2}r)$ also in $^{46}$V, because the relative energy
 of the lowest $T=0$ and $T=1$ states in odd-odd nuclei is sensitive
 to these parameters.  If we enlarge the absolute values of
 $\Delta k^0 (f_{7/2}f_{7/2})$ and $\Delta k^0(f_{7/2}r)$, we get
 better energies for $T=0$ low-spin states but too high energies
 for $T=0$ high-spin states in $^{46}$V.  The same tendency is seen
 in the column D of Fig. \ref{fig2}, where the larger the values of
 $\Delta k^0 (f_{7/2}f_{7/2})$ and $\Delta k^0(f_{7/2}r)$ become,
 the higher the high-spin levels go up.  Figures \ref{fig1} and
 \ref{fig2} say that the parameters $\Delta k^1 (f_{7/2}f_{7/2})$,
 $\Delta k^1(f_{7/2}r)$, $\Delta k^0 (f_{7/2}f_{7/2})$ and
 $\Delta k^0(f_{7/2}r)$ must not change too much from the values
 in Eqs. (\ref{eq:7}) and (\ref{eq:9}).

  Using the parameter set D, we obtain the ground-state energies shown
 in Fig. \ref{fig3}, where the obtained result is compared with the
 result of (I) (cal A) and with the experimental ground-state energies
 (see (I)).  The modification in this paper improves
 the binding energies of nuclei with large $T$ such as the Ca isotopes,
 as compared with the original model.  Figure \ref{fig3}, however,
 tells slight overbinding of the $N \approx Z$ nuclei with small $T$.
  For instance, the model (denoted by +) lowers a little the excitation
 energies of the $T=1$ and $T=2$ isobaric analogue states in $^{48}$Cr,
 while the original model (denoted by $\ast$) reproduces them.
 We keep the fixed values (\ref{eq:7}) and (\ref{eq:9}) of
 the parameters $\Delta k^1$ and $\Delta k^0$ for all the $A$=42-51
 nuclei in Fig. \ref{fig3}, but the figure suggests
 that the additional monopole terms are too strong for the $A$=50 and
 $A$=51 nuclei with large $T$.  Possibly the parameters $\Delta k^1$
 and $\Delta k^0$ should have $A$ dependence.
  The present modification still leaves a room for fine improvement.
 Our concern is, however, to sketch out functions of various typical
 interactions.  The guiding principle in this paper is to cure the 
 above-mentioned two flaws of the original model by using as few
 number of parameters as possible.  We are content with the results in
 Figs. \ref{fig1}, \ref{fig2} and \ref{fig3}.  The modified model
 with the parameter set D fairly well reproduces
 the binding energies and level schemes of $^{48}$Ca and $^{47}$Ti.
 We shall examine this modified model in the following sections.

%===============================================================
\section{Effects of the additional monopole terms on even-$A$ nuclei}

   Let us see the effects of the additional monopole terms on even-A
 nuclei, to which the original model without these terms is applied
 in (I).
 
   The original $P_0+P_2+QQ+V^0_{\pi \nu}$ interaction well describes
 the collective yrast states of $N \approx Z$ even-even nuclei.
 The additional monopole terms with the parameters (\ref{eq:9}) have
 little effect on these nuclei, though we do not show figures of
 the results except $^{50}$Cr.  The discussions about the collective
 yrast states of $^{46}$Ti, $^{48}$Cr and $^{50}$Cr in (I) remain valid
 also in the present modified model.  The level structure including
 non-yrast levels and $B(E2)$ in the ground-state band are 
 hardly changed in $^{46}$Ti and $^{48}$Cr.  The backbending plot of
 the $^{48}$Cr rotational band is fairly well and the quality is
 the same as that in (I).  Speaking strictly, the additional monopole
 terms do not improve the excitation energies of $8^+_1$, $10^+_1$
 and $12^+_1$ in the backbending region. The somewhat awkward change
 from the most collective states ($0^+_1 - 8^+_1$) to the high-spin
 states in the calculated ground-state band of $^{48}$Cr seems
 to originate in the $P_0+P_2+QQ$ force.
 
%===  Fig. 4 ========================================================
\begin{figure}
\begin{center}
    \epsfig{file=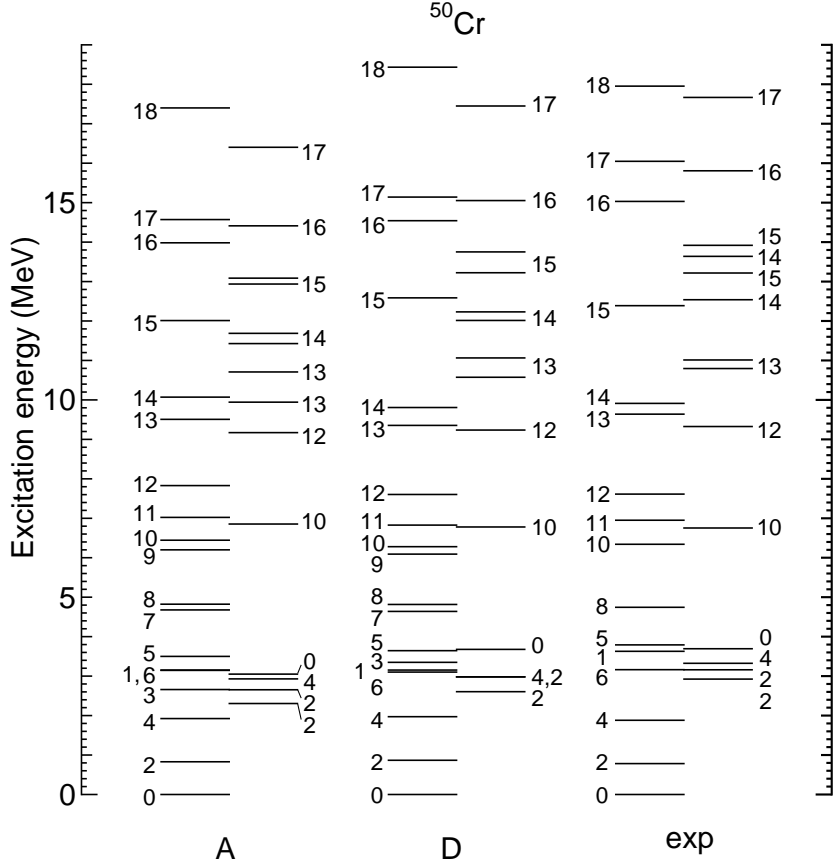,width=0.8\textwidth}
\caption{Energy levels of $^{50}$Cr calculated with the original and
         modified models (A and D), compared with the observed levels.}
\label{fig4}
\end{center}
\end{figure}
%====================================================================

  The modified model (D) produces a little better level structure
 than the original model (A) in $^{50}$Cr, as shown in Fig. \ref{fig4}.
 Not only the yrast levels but also other levels correspond better to
 the observed levels.  The low-spin states $2^+_2$, $4^+_2$, $2^+_3$
 and $0^+_2$, and the high-spin states $13^+_2$, $13^+_3$, $14^+_2$,
 $14^+_3$, $15^+_2$, $15^+_3$, $16^+_2$ and $17^+_2$ come nearer to
 the observed positions.
 In $^{50}$Cr, however, the $B(E2)$ values between the most collective
 yrast states are considerably reduced: $B(E2:J_1 \rightarrow (J-2)_1)$
 in e$^2$fm$^4$ is 197 for $J=2$; 277 for $J=4$; 122 for $J=6$; 121 for
 $J=8$ in the calculation D, while it is 215 for $J=2$; 302 for $J=4$;
 211 for $J=6$; 200 for $J=8$ in the calculation A.  The modification
 of the model has a tendency to push high-spin yrast states upward,
 as seen in Fig. \ref{fig4}.  The same tendency is observed
 in the results of $^{46}$Ti and $^{48}$Cr.  

%===  Fig. 5 ========================================================
\begin{figure}
\begin{center}
    \epsfig{file=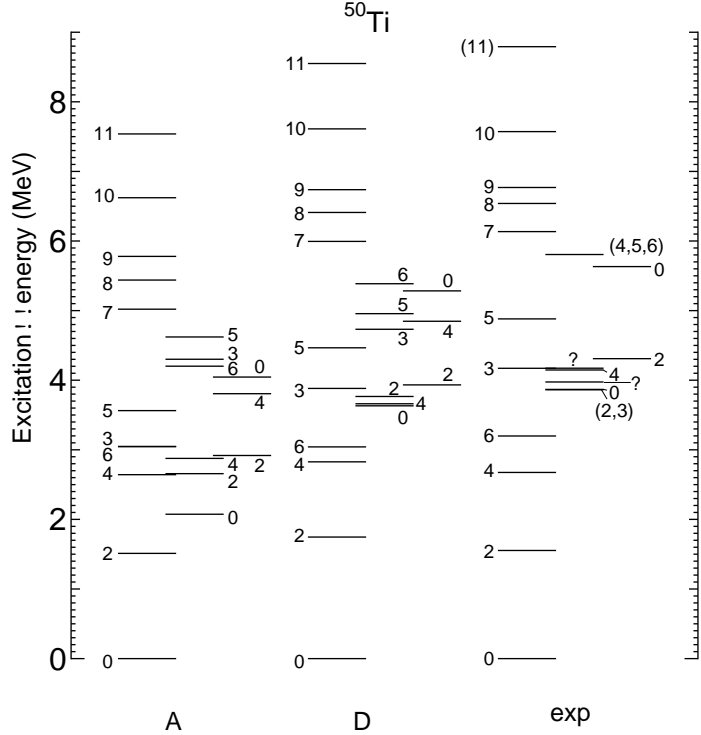,width=0.67\textwidth}
\caption{Energy levels of $^{50}$Ti calculated with the original and
         modified models (A and D), compared with the observed levels.}
\label{fig5}
\end{center}
\end{figure}
%====================================================================

  The modified model works clearly better as compared with the original
 model, when $N$ separates from $Z$, as in $^{48}$Ca.  We show the
 level scheme of $^{50}$Ti as a good example in Fig. \ref{fig5},
 where yrast levels and other low-lying levels obtained by the original
 model (A) and modified one (D) are compared with the observed levels.
 The additional monopole terms move the yrast levels above $5^+_1$ very
 nearer to the observed ones, and push the non-yrast levels upward,
 laying them at good positions.

%===  Fig. 6 ========================================================
\begin{figure}
\begin{center}
    \epsfig{file=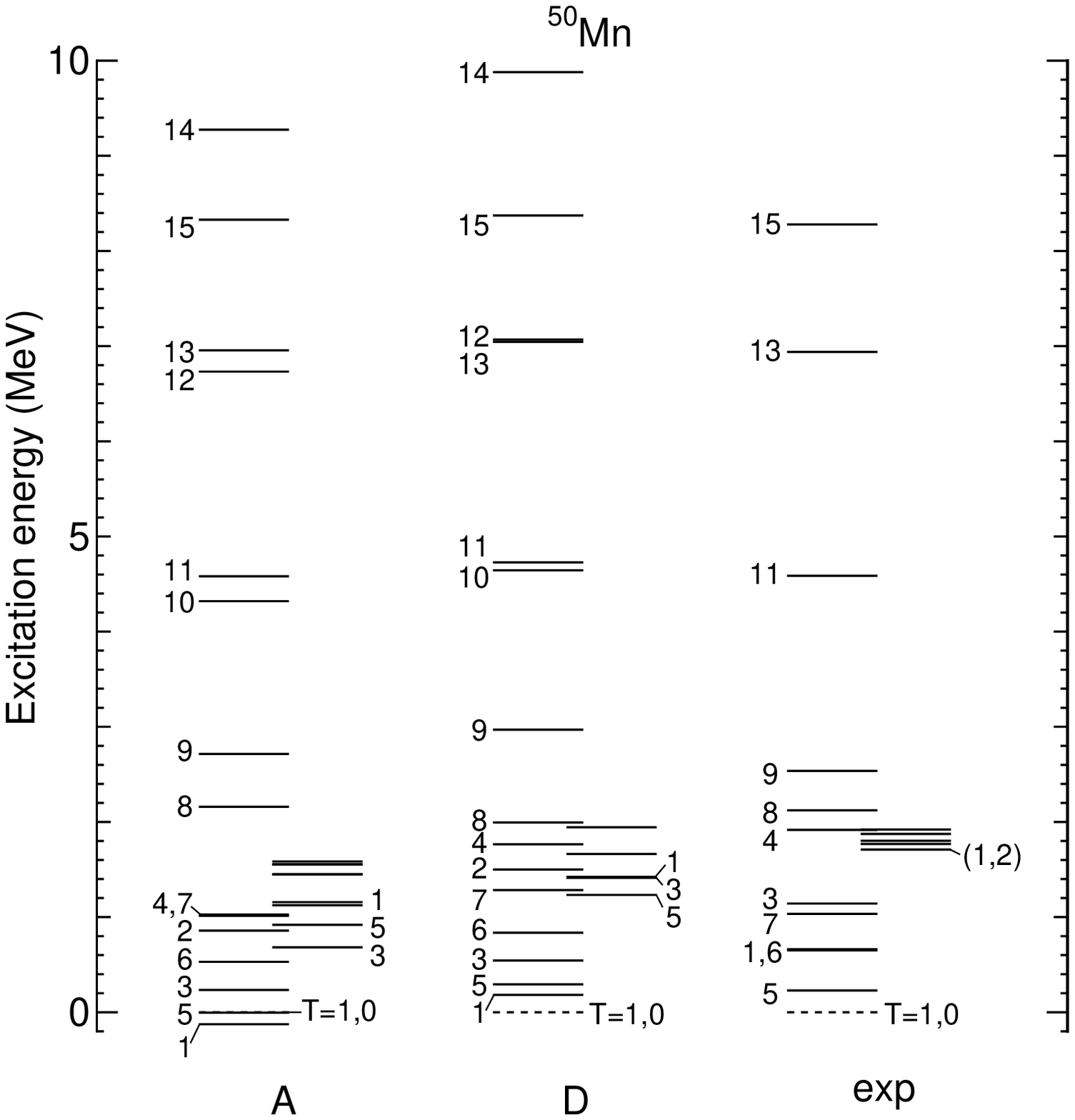,width=0.67\textwidth}
\caption{Energy levels of $^{50}$Mn calculated with the original and
         modified models (A and D), compared with the observed levels.}
\label{fig6}
\end{center}
\end{figure}
%====================================================================

%===  Fig. 7 ========================================================
\begin{figure}
\begin{center}
    \epsfig{file=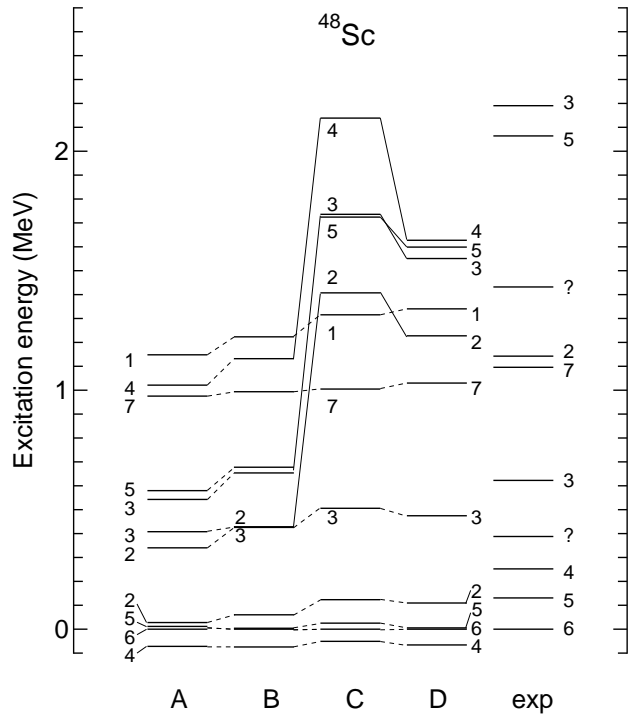,width=0.60\textwidth}
\caption{Energy levels of $^{48}$Sc calculated with four different sets
         of parameters, compared with the observed levels.}
\label{fig7}
\end{center}
\end{figure}
%====================================================================

  Next, we examine the applicability of our modified model in odd-odd
 nuclei which are most sensitive to the interaction employed.  
 The previous paper (I) has shown that the $P_0+P_2+QQ+V^0_{\pi \nu}$
 interaction can describe the yrast level structure in $^{46}$V, 
 $^{48}$V and $^{50}$Mn beyond expectation.  The introduction of the 
 additional monopole terms reproduces relatively better results
 in these nuclei.  In $^{46}$V, for instance, the lowest $3^+$ state
 with $T=0$ lies at 0.24 MeV in the original model A, while it lies
 at 0.59 MeV, coming near to the observed excitation energy 0.80 MeV,
 in the modified model D.  The yrast and non-yrast levels correspond
 better to the observed ones in the present modified model,
 though the high-spin state $15^+_1$ goes up a little.

  As shown in Fig. \ref{fig6}, the relative positions of the $T=0$ and
 $T=1$ states are improved also in $^{50}$Mn.  The original model (A)
 lays the lowest $T=0$ state with $J^\pi=1^+$ below the $T=1$, $0^+_1$
 state, but the modified model (D) correctly yields the ground state
 with $T=1$, $J^\pi=0^+$.  If we regard the $T=0$ yrast levels with 
 $J \geq 5$ as a collective band on the band-head state $5^+_1$, the
 band is well reproduced by our model.  The additional monopole terms
 push non-yrast (noncollective) levels toward the observed ones.  The
 yrast states with smaller spins ($J<5$) than the band-head state,
 however, remain at low energy, and hence the order of low-lying levels
 does not correspond to the experimental order.
 
   The same problem is observed in odd-odd nuclei away from $N=Z$,
 though the introduction of the additional monopole terms is very
 effective in improving the level scheme.  The modified model well
 describes the collective yrast band, but cannot reproduce the correct
 order of low-lying levels and fails to give the correct ground state
 in $^{48}$V and $^{48}$Sc.  In $^{48}$V, the lowest state $2^+_1$ is
 at -0.20 MeV below the ground state $4^+_1$ in the calculation D.
 The difference between the calculated and observed yrast levels is
 0.5 MeV at most.
   We show another example in Fig. \ref{fig7}, where results of the
 calculations A, B, C and D are compared with the observed level scheme
 of $^{48}$Sc. Note here that Fig. \ref{fig7} is drawn in a small-scale
 as compared with the previous figures.  The additional monopole
 terms produce very improved results.  The correspondences of the
 excitation energies and level density between theory and experiment
 become better in the modified model D. Although the calculation cannot
 reproduce the correct ground state, the discrepancy does not exceed
 0.4 MeV.
 The results in odd-odd nuclei suggest that fine improvement would be
 possible in our extended $P+QQ$ model.

%===============================================================
\section{Results in the cross-conjugate nuclei
  $^{47}$T\lowercase{i} - $^{49}$V and $^{47}$V - $^{49}$Cr}

  The accumulation of experimental and theoretical studies has been
 clarifying the structure of $A$=47 and $A$=49 nuclei.
 Mart\'{\i}nez-Pinedo {\it et al.} \cite{Martinez2,Poves2} made
 an exhaustive study of these nuclei by the full $fp$ shell model
 calculation using the KB3 interaction.  These nuclei are very suitable
 for examining the quality of our model.  We direct our attention
 to the pairs of nuclei $^{47}$Ti - $^{49}$V and $^{47}$V - $^{49}$Cr
 which have the cross-conjugate configurations ($(f_{7/2})^{7p}$ and
 $(f_{7/2})^{7h}$) in the single $j$ limit. 
 The energy levels of $^{47}$V and $^{47}$Cr ($^{49}$Cr and $^{49}$Mn)
 are equivalent in our model because of the isospin invariant
 Hamiltonian.  We discuss the structure of $^{47}$V and $^{49}$Cr
 for simplicity.  Only negative-parity states of these nuclei are
 considered in the model space.  We use the effective charge
 $e_{eff}=0.5e$ and the harmonic-oscillator range parameter
 $b_0=1.01 A^{1/6}$ fm in the calculations.  These values are the same
 as those in Refs. \cite{Martinez2,Poves2}.

\subsection{$^{47}$T\lowercase{i} - $^{49}$V}

   We have already seen in Fig. 2 that the extended $P+QQ$ model
 satisfactorily reproduces the yrast levels of $^{47}$Ti.  This is
 clearer in the spin-energy graph in Fig. 8, where our model (cal D)
 correctly traces the observed staggering gait depending on the spins
 $(2J \pm 1)/2$.  The result of our extended $P+QQ$ model is almost
 comparable to that of Mart\'{\i}nez-Pinedo {\it et al.} with the KB3
 interaction.  The KB3 interaction reproduces better the $17/2^-_1$,
 $19/2^-_1$ and $21/2^-_1$ levels than ours, while our interaction
 correctly reproduces the order of $9/2^-_1$ and $11/2^-_1$ and the
 position of the high-spin level $27/2^-_1$ as compared with the KB3.
 The insufficiency of our model with respect to the energy of
 $21/2^-_1$ is possibly related to the remaining flaw around
 $8^+_1 - 12^+_1$ in $^{48}$Cr mentioned in the previous section.
 The staggering in the spin-energy graph of Fig. 8 reminds us of that
 in the odd-odd nuclei $^{46}$V, $^{48}$V and $^{50}$Mn discussed
 in (I).  There must be a similar mechanism of excitation (rotation)
 in the yrast bands of $^{47}$Ti and $^{48}$V.

%===  Fig. 8 ========================================================
\begin{figure}
\begin{center}
    \epsfig{file=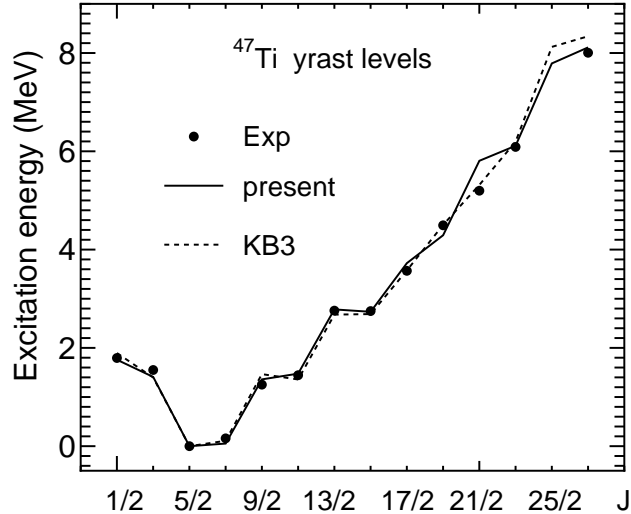,width=0.6\textwidth}
\caption{Theoretical and experimental spin-energy relation
    in the yrast band of $^{47}$Ti. The result of the full $fp$ shell
    model calculation with the KB3 interaction \cite{Martinez2}
    is also shown by the dotted line.}
\label{fig8}
\end{center}
\end{figure}
%====================================================================

  Table \ref{table1} shows electric quadrupole properties of the 
 yrast and other states in $^{47}$Ti.  The calculated $B(E2)$ values 
 correspond to the tendency of the observed ones \cite{Burrows1}
 to some extent, but the quality of the data may not be sufficient
 \cite{Martinez2}.
  We calculated the spectroscopic quadrupole moment $Q_{spec}$ and
 intrinsic quadrupole moments $Q^{(s)}_0$ and $Q^{(t)}_0$ for the yrast
 states in addition to $B(E2)$. (We use the same notations $Q^{(s)}_0$
 and $Q^{(t)}_0$ as Mart\'{\i}nez-Pinedo {\it et al.}
 \cite{Martinez2}.)  With respect to these electric quadrupole
 properties, our predictions in the smaller model space
 $(f_{7/2}, p_{3/2}, p_{1/2})$ are very similar to those of the full
 $fp$ shell model with the KB3 interaction.

%===============  Ti(47)  ========================
\begin{footnotesize}
\begin{table}
\caption{Electric quadrupole properties of the yrast and other states
 in $^{47}$Ti: ${B(E2)}$ in $e^2$ fm$^4$; $Q_{spec}$, $Q^{(s)}_0$ and
 $Q^{(t)}_0$ in $e$ fm$^2$. The present results are compared with
 the experimental ones and those of Mart\'{\i}nez-Pinedo et al.}
\label{table1}
\begin{center}
\begin{tabular}{rcccccccc} \hline
   & Expt. & \multicolumn{4}{c}{present}
   & \multicolumn{3}{c}{Mart\'{\i}nez-Pinedo} \\
 $J_n \rightarrow J^\prime_m$ & $B(E2)$
 & $B(E2)$ & $Q_{spec}$ & $Q^{(s)}_0$ & $Q^{(t)}_0$
 & $B(E2)$ & $Q^{(s)}_0$ & $Q^{(t)}_0$ \\ \hline
   $5/2_1$ \ \ \ \ \ &  
  &     & 23.6 & 66 &      &     &  63.4 &      \\
 $7/2_1 \rightarrow 5/2_1$ & 252(50)
  & 145 &  6.4 & 96 & 64   & 140 & 120.3 & 61.6 \\
 $9/2_1 \rightarrow 5/2_1$ &  70(30)
  &  56 & -5.5 & 61 & 75   &  55 &  44.1 & 72.7 \\
       $\rightarrow 7/2_1$ & 191(40)
  & 136 &      &    & 67   & 102 &       & 57.1 \\
 $11/2_1 \rightarrow 7/2_1$ & 159(25)
  &  98 & -6.8 & 36 & 76   &  98 &  10.9 & 74.7 \\
        $\rightarrow 9/2_1$ & 705(605)
  &  92 &      &    & 63   &  83 &       & 58.6 \\
 $13/2_1 \rightarrow 9/2_1$ & 
  & 109 & -14.3 & 57 & 71   &    &  60.7 & 67.5 \\
        $\rightarrow 11/2_1$ & 
  &  68 &       &    & 62   &    &       & 57.3 \\
 $15/2_1 \rightarrow 11/2_1$ & 135(26)
  & 102 & -11.5 & 39 & 64   & 111 & 29.8 & 66.0 \\
        $\rightarrow 13/2_1$ & 
  &  47 &       &    & 58   &    &       & 50.2 \\
 $17/2_1 \rightarrow 13/2_1$ & 
  &  89 & -19.6 & 60 & 58   &    &  69.7 & 59.7 \\
        $\rightarrow 15/2_1$ & 604$^{+70}_{-640}$
  &  34 &       &    & 55   & 41 &       & 59.3 \\
 $19/2_1 \rightarrow 15/2_1$ & 
  &  57 & -19.4 & 55 & 45   &    &  45.6 & 51.9 \\
        $\rightarrow 17/2_1$ & $<50$
  &  30 &       &    & 57   & 25 &       & 50.6 \\
 $21/2_1 \rightarrow 17/2_1$ & 
  &  41 & -21.1 & 57 & 37   &    &  66.6 & 40.4 \\
        $\rightarrow 19/2_1$ &
  &   7 &       &    & 30   &    &       & 40.3 \\
 $23/2_1 \rightarrow 19/2_1$ & 
  &  43 & -19.6 & 51 & 38   &    &  53.5 & 43.1 \\
        $\rightarrow 21/2_1$ &
  &  21 &       &    & 56   &    &       & 56.8 \\
 $25/2_1 \rightarrow 21/2_1$ & 
  &  12 & -10.2 & 26 & 19   &    &  30.1 & 20.3 \\
        $\rightarrow 23/2_1$ &
  & 1.4 &       &    & 16   &    &       &  9.1 \\
 $27/2_1 \rightarrow 23/2_1$ & 
  &  25 & -17.5 & 43 & 28   &    &  46.7 & 28.8 \\
        $\rightarrow 25/2_1$ &
  &  19 &       &    & 62   &    &       & 65.8 \\
 $3/2_1 \rightarrow 5/2_1$ & 3.3(15)
  &  9.9 &   &   &     & 22 &   &  \\
       $\rightarrow 7/2_1$ & 39(11)
  &  1.4 &   &   &     & 46 &   &  \\
 $1/2_1 \rightarrow 5/2_1$ & $<17$
  &  11  &   &   &     & 21 &   &  \\
 $3/2_2 \rightarrow 5/2_1$ & $<272$
  &  9.4 &   &   &     & 0.48 &   &  \\
       $\rightarrow 7/2_1$ & 36.3(70)
  &  60  &   &   &     & 8.3  &   &  \\ \hline
\end{tabular}
\end{center}
\end{table}
\end{footnotesize}
%=========================================================
 
  For the ground state $5/2^-_1$, The calculated spectroscopic
 quadrupole moment $Q_{spec}=23.6$ $e$ fm$^2$ is nearly equal to
 the value 22.7 $e$ fm$^2$ of Ref. \cite{Martinez2} but does not well
 reproduce the observed value 30.3 $e$ fm$^2$.
  The calculated results show that $B(E2:J \rightarrow J-1)$ is larger
 than $B(E2:J \rightarrow J-2)$ up to $J=9/2$, while 
 $B(E2:J \rightarrow J-2)$ exceeds $B(E2:J \rightarrow J-1)$ above 
 $J=9/2$ in the yrast band.  The full $fp$ shell model with the KB3
 interaction seems to give a similar prediction for the $B(E2)$ ratio
 from the values of $Q^{(t)}_0$. The observed $\gamma$ transitions
 \cite{Cameron1} are consistent with the theoretical prediction
 up to $15/2^-_1$ but show a different feature from the prediction
 for $J \geq 1 7/2^-_1$.
 In the calculation, $Q_{spec}$ changes the value considerably
 at $17/2^-_1$ and above it the $B(E2)$ values become smaller.
 Our model gives different $B(E2)$ values to the $3/2^-_1$ and
 $1/2^-_1$ states from Ref. \cite{Martinez2}.

%===  Fig. 9 ========================================================
\begin{figure}[b]
\begin{center}
    \epsfig{file=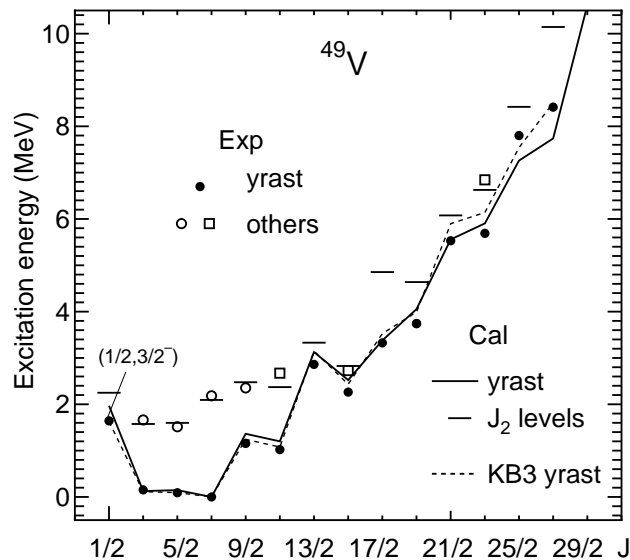,width=0.6\textwidth}
\caption{Calculated and observed energy levels of $^{49}$V.
         The lowest two levels of each $J$ (yrast and $J_2$) obtained
         by the present model are compared with the observed ones.
         The yrast levels obtained by the full $fp$ shell model
         calculation with the KB3 interaction \cite{Martinez2}
         are shown by the dotted line.}
\label{fig9}
\end{center}
\end{figure}
%====================================================================

  Energy levels of $^{49}_{23}$V$_{26}$ obtained by the present model
 are compared with observed levels \cite{ToI,Cameron} in Fig.
 \ref{fig9}.
 Our model reproduces not only the yrast levels but also the
 second lowest levels from $J^\pi=3/2^-$ to $11/2^-$, and nearly traces
 the staggering gait in the spin-energy graph of the yrast states.
 For the yrast levels, the correspondence between theory and experiment
 in $^{49}$V is somewhat worse than that in $^{47}$Ti at high spin.
 It is, however, notable that the two spin-energy graphs of the
 cross-conjugate nuclei $^{47}$Ti and $^{49}$V in Figs. \ref{fig8} and
 \ref{fig9} resemble each other and are well described by our model.

%===================  V(49)  ====================
\begin{footnotesize}
\begin{table}
\caption{Electric quadrupole properties of the yrast and other states
 in $^{49}$V: ${B(E2)}$ in $e^2$ fm$^4$; $Q_{spec}$ in $e$ fm$^2$.}
\label{table2}
\begin{center}
\begin{tabular}{rccccc} \hline
   & Expt. & \multicolumn{2}{c}{present}
   & \multicolumn{2}{c}{Mart\'{\i}nez-Pinedo} \\
 $J_n \rightarrow J^\prime_m$ & $B(E2)$
 & $B(E2)$ & $Q_{spec}$ & $B(E2)$ & $Q_{spec}$ \\ \hline
   $7/2_1$ \ \ \ \ \         &         &     & -12.2 &     & -11.1 \\
 $5/2_1 \rightarrow 7/2_1$   &         & 246 &  -7.3 &     &       \\
 $3/2_1 \rightarrow 5/2_1$   &         & 355 &  19.5 &     &       \\
       $\rightarrow 7/2_1$   & 204(6)  & 205 &       & 196 &       \\
 $11/2_1 \rightarrow 7/2_1$  & 149(27) & 158 & -20.7 & 157 &       \\
 $9/2_1 \rightarrow 5/2_1$   &  84(24) & 116 & -25.4 &  88 &       \\
       $\rightarrow 7/2_1$   &  63(19) &  70 &       &  41 &       \\
       $\rightarrow 11/2_1$  &         &  98 &       &     &       \\
 $15/2_1 \rightarrow 11/2_1$ & 298$^{+85}_{-180}$
                                       & 107 &  -2.7 & 140 &       \\
 $13/2_1 \rightarrow 9/2_1$  & 290(200)& 140 & -26.3 & 110 &       \\
        $\rightarrow 11/2_1$ &         &  24 &       &     &       \\
        $\rightarrow 15/2_1$ &         &  20 &       &     &       \\
       \hline \hline
       \multicolumn{6}{c}{present work} \\
  $J_1$ & $Q_{spec}$ & $\rightarrow J^\prime_1$  & $B(E2)$
   &  $\rightarrow J^\prime_1$ &  $B(E2)$  \\ \hline
 $17/2_1$ & 29.3
          & $\rightarrow 13/2_1$ & 0.9 & $\rightarrow 15/2_1$ & 33 \\
 $19/2_1$ &  7.1
          & $\rightarrow 15/2_1$ &  13 & $\rightarrow 17/2_1$ & 79 \\
 $21/2_1$ & 10.6
          & $\rightarrow 17/2_1$ &  52 & $\rightarrow 19/2_1$ & 42 \\
 $23/2_1$ &  0.9
          & $\rightarrow 19/2_1$ &  68 & $\rightarrow 21/2_1$ & 55 \\
 $25/2_1$ & -2.9
          & $\rightarrow 21/2_1$ &  44 & $\rightarrow 23/2_1$ & 37 \\
 $27/2_1$ & -6.5
          & $\rightarrow 23/2_1$ &  43 & $\rightarrow 25/2_1$ & 24 \\
   \hline
\end{tabular}
\end{center}
\end{table}
\end{footnotesize}
%=========================================================

   Table \ref{table2} shows calculated $B(E2)$ values and observed ones
 \cite{Burrows2}, and calculated values of the spectroscopic quadrupole
 moment $Q_{spec}$.  The results of Mart\'{\i}nez-Pinedo {\it et al.}
 \cite{Martinez2} are also shown for comparison.  The $B(E2)$ values
 obtained by our model are comparable with those of Ref.
 \cite{Martinez2}, and roughly consistent with the observed $B(E2)$
 values though the data are of poor quality.  The spectroscopic
 quadrupole moment of the ground state $7/2^-_1$ obtained by our model
 is nearly equal to that of Ref. \cite{Martinez2}.  The calculated
 values of $Q_{spec}$ say that the yrast band changes the structure at
 the $17/2^-_1$ state.  Correspondingly, the absolute $B(E2)$ values
 and the ratio
  $B(E2:J_1 \rightarrow (J-2)_1)/B(E2:J_1 \rightarrow (J-1)_1)$
 show different features before and after $17/2^-_1$.
 This structure change appears in the spin-energy graph of Fig.
 \ref{fig9}, where the staggering is large from $9/2^-_1$ to $11/2^-_1$
 and from $13/2^-_1$ to $15/2^-_1$, and then becomes gentle above
 the $15/2^-_1$ state, both in theory and experiment.

\subsection{$^{47}$V - $^{49}$Cr}

\begin{figure}
\begin{center}
    \epsfig{file=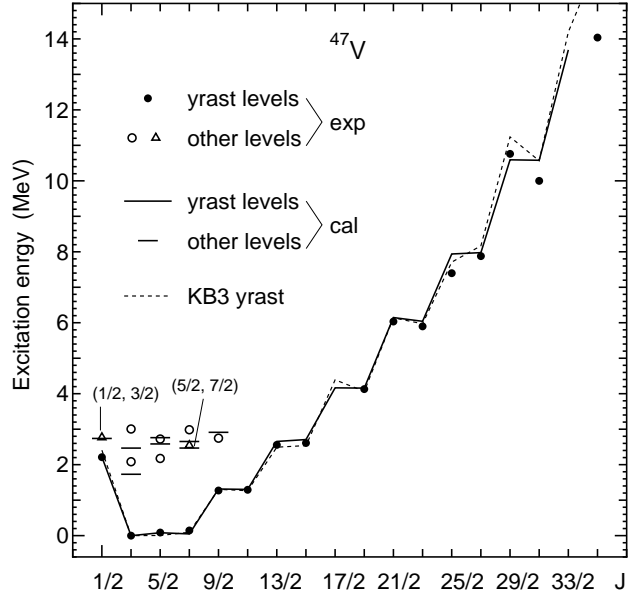,width=0.6\textwidth}
\caption{Calculated and observed energy levels of $^{47}$V.
         The yrast levels obtained by the full $fp$ shell model
         calculation with the KB3 interaction \cite{Martinez2} are
         shown by the dotted line.}
\label{fig10}
\end{center}
\end{figure}

   Figure \ref{fig10} shows calculated energy levels and observed ones 
 \cite{ToI,Cameron1,Cameron2,Bentley} of $^{47}_{23}$V$_{24}$: the
 yrast levels and other low-lying levels.
  The experimental energy of the $21/2^-_1$ state shown in Fig.
 \ref{fig10} is chosen from Ref. \cite{Bentley}, which is different
 from that of Ref. \cite{Cameron2}.
  The present model excellently reproduces the yrast levels up to the
 $23/2^-$ state.  The correspondence between theory and experiment
 with respect to other low-lying states is also considerably well.
  Our model reproduces better the high-spin levels $27/2^-_1$ and
 $29/2^-_1$ than the full $fp$ shell model with the KB3 interaction
 \cite{Martinez2,Poves2}, and correctly yields the regular order of
 the pair levels ($25/2^-_1$, $27/2^-_1$) and the inverse order of
 ($31/2^-_1$, $29/2^-_1$).  However, the observed staggering gait from
 $25/2^-_1$ to $31/2^-_1$ in the spin-energy graph is not well
 reproduced in contrast to the success of Refs.
 \cite{Martinez2,Poves2}.
 This is a question to our model which is expected to describe
 the collective band well.  We shall touch upon this question
 in the next subsection.

%===  Fig. 10 ========================================================
%====================================================================

%===================  V(47)  ====================
\begin{footnotesize}
\begin{table}
\caption{Electric quadrupole properties in $^{47}$V: ${B(E2)}$ in
 $e^2$ fm$^4$; $Q_{spec}$ and $Q_0$ in $e$ fm$^2$.}
\label{table3}
\begin{center}
\begin{tabular}{rcccccccc} \hline
   & Expt. & \multicolumn{4}{c}{present}
   & \multicolumn{3}{c}{Mart\'{\i}nez-Pinedo} \\
 $J_n \rightarrow J^\prime_m$ & $B(E2)$
 & $B(E2)$ & $Q_{spec}$ & $Q^{(s)}_0$ & $Q^{(t)}_0$
 & $B(E2)$ & $Q^{(s)}_0$ & $Q^{(t)}_0$ \\ \hline
   $3/2_1$ \ \ \ \ \ &  
  &     &  19.7 &  99 &       &     &  100 &    \\
 $5/2_1 \rightarrow 3/2_1$ & 
  & 266 &  -7.4 & 104 &  88   & 251 &  138 & 87 \\
 $7/2_1 \rightarrow 3/2_1$ &  74(10)
  & 105 & -13.2 &  66 &  86   & 106 &   67 & 88 \\
       $\rightarrow 5/2_1$ &
  & 192 &       &     &  95   &     &      & 99 \\
 $9/2_1 \rightarrow 5/2_1$ & 160(50)
  & 137 & -25.8 &  94 &  80   & 138 &  101 & 82 \\
       $\rightarrow 7/2_1$ &
  &  92 &       &     &  81   &  76 &      & 75 \\
 $11/2_1 \rightarrow 7/2_1$ & 200(100)
  & 175 & -23.1 &  72 &  83   & 186 &   69 & 87 \\
        $\rightarrow 9/2_1$ & 
  &  87 &       &     &  94   &     &      & 100 \\
 $13/2_1 \rightarrow 9/2_1$ & 
  & 167 & -31.9 &  91 &  78   &     &  101 & 77 \\
        $\rightarrow 11/2_1$ & 
  &  36 &       &     &  71   &     &      & 66 \\
 $15/2_1 \rightarrow 11/2_1$ & $<79$ ($>97$)
  & 182 & -26.5 &  71 &  78   &     &   69 & 81 \\
        $\rightarrow 13/2_1$ & 
  &  42 &       &     &  88   &     &      & 103 \\
 $17/2_1 \rightarrow 13/2_1$ & 
  &  91 & -13.3 &  34 &  54   &     &  -5.5 & 17 \\
        $\rightarrow 15/2_1$ & 
  & 4.8 &       &     &  33   &     &       & 22 \\
 $19/2_1 \rightarrow 15/2_1$ & 140(40)
  & 112 & -16.1 &  40 &  59   &     &    39 & 65 \\
        $\rightarrow 17/2_1$ & 
  & 2.7 &       &     & 28   &     &       & 106 \\
 $21/2_1 \rightarrow 17/2_1$ & 
  & 112 & -16.1 &  40 &  59   &     &    38 & 54 \\
        $\rightarrow 19/2_1$ &
  & 4.0 &       &     &  37   &     &       & 35 \\
 $23/2_1 \rightarrow 19/2_1$ & 93(14)
  & 137 & -18.2 &  43 &  64   &     &    42 & 66 \\
        $\rightarrow 21/2_1$ &
  & 4.3 &       &     &  42   &     &       & 43 \\
 $25/2_1 \rightarrow 21/2_1$ & 10(3)
  &  79 & -15.4 &  36 &  49   &     &    45 & 57 \\
        $\rightarrow 23/2_1$ &
  & 8.7 &       &     &  65   &     &       & 53 \\
 $27/2_1 \rightarrow 23/2_1$ & 117(28)
  &  99 & -17.9 &  41 &  54   &     &    38 & 55 \\
        $\rightarrow 25/2_1$ &
  & 7.8 &       &     &  67   &     &       & 63 \\
 $29/2_1 \rightarrow 25/2_1$ & 
  &  32 & -15.4 &  35 &  31   &     &    35 & 32 \\
        $\rightarrow 27/2_1$ &
  & 0.1 &       &     & 7.1   &     &       & 22 \\
 $31/2_1 \rightarrow 27/2_1$ & 55(6)
  &  44 & -18.5 &  42 &  36   &     &    42 & 41 \\
        $\rightarrow 29/2_1$ &
  & 3.6 &       &     &  52   &     &       & 69 \\
 $17/2_2 \rightarrow 13/2_1$ & 
  &  55 &  16.5 & -42 &  42   &     &    68 & 66 \\
%        $\rightarrow 15/2_1$ & 
%  & 8.4 &       &     &  44   &     &       & 41 \\
    \hline
\end{tabular}
\end{center}
\end{table}
\end{footnotesize}
%=========================================================

  In Table \ref{table3}, calculated electric quadrupole properties
 $B(E2)$, $Q_{spec}$, $Q^{(s)}_0$ and $Q^{(t)}_0$ are compared with
 those of Ref. \cite{Martinez2} and with the observed $B(E2)$ values.
  Our results, which are very similar to those of the full $fp$ shell
 model calculation with the KB3 interaction except the $17/2^-_1$ and
 $17/2^-_2$ states, correspond well to the observed $B(E2)$ values
 except $B(E2:25/2^-_1 \rightarrow 21/2^-_1)$.
 The calculated large values of $B(E2:7/2^-_1 \rightarrow 5/2^-_1)$ 
 and $B(E2:5/2^-_1 \rightarrow 3/2^-_1)$ explain the observed strong
 $\gamma$ transitions $7/2^-_1 \rightarrow 5/2^-_1 \rightarrow 3/2^-_1$
 \cite{Cameron2}, and the large values of
 $B(E2:J^-_1 \rightarrow (J-2)^-_1)$ when $J=11/2$, 15/2, 19/2 and 23/2
 are consistent with the observed cascade transitions
 $23/2^-_1 \rightarrow 19/2^-_1 \rightarrow 15/2^-_1 \rightarrow
 11/2^-_1 \rightarrow 7/2^-_1$ \cite{Cameron2,Bentley}.
  Other observed $\gamma$ transitions for the high-spin states are
 not necessarily understood by our result that
 $B(E2:J^-_1 \rightarrow (J-2)^-_1)$ is much larger than
 $B(E2:J^-_1 \rightarrow (J-1)^-_1)$ above $7/2^-_1$.
 The calculated spectroscopic quadrupole moment $Q_{spec}$ suggests
 a change in the structure of the yrast band at $17/2^-_1$, which is
 consistent with the change appearing near $17/2^-_1$ in the Coulomb
 energy difference between the mirror nuclei $^{47}$V and $^{47}$Cr
 \cite{Bentley}.  
 Our model and the full $fp$ shell model with the KB3
 interaction give different $Q^{(s)}_0$ and $Q^{(t)}_0$ for the
 $17/2^-_1$ and $17/2^-_2$ states.  The two states in the latter model
 (where the two are close by) seem to be in reverse order to those
 in our model.  Experimental data about the $17/2^-$ states
 are unfortunately missing for discussing the difference.
 This is possibly due to the degeneracy of the $17/2^-_1$ and
 $19/2^-_1$ states predicted by our model.
 The value of $Q_{spec}$ shows a moderate change for the
 $17/2^-_1$, $19/2^-_1$, $23/2^-_1$, $27/2 ^-_1$ and $31/2^-_1$ states
 after the structure change (maybe band crossing) at $17/2^-_1$ in our
 model.

 %===  Fig. 11 ========================================================
\begin{figure}[b]
\begin{center}
    \epsfig{file=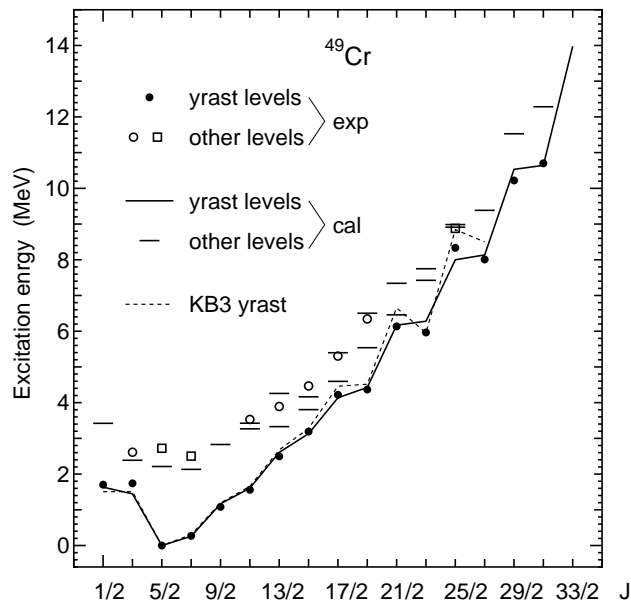,width=0.6\textwidth}
\caption{Calculated and observed energy levels of $^{49}$Cr.
         The yrast levels obtained by the full $fp$ shell model
         calculation with the KB3 interaction \cite{Martinez2} are
         shown by the dotted line.}
\label{fig11}
\end{center}
\end{figure}
%====================================================================

%===================  Cr(49)  ====================
\begin{footnotesize}
\begin{table}
\caption{Electric quadrupole properties in $^{49}$Cr:
      ${B(E2)}$ in $e^2$ fm$^4$; $Q_{spec}$ and $Q_0$ in $e$ fm$^2$.}
\label{table4}
\begin{center}
\begin{tabular}{rcccccccc} \hline
   & Expt. & \multicolumn{4}{c}{present}
   & \multicolumn{3}{c}{Mart\'{\i}nez-Pinedo} \\
 $J_n \rightarrow J^\prime_m$ & $B(E2)$
 & $B(E2)$ & $Q_{spec}$ & $Q^{(s)}_0$ & $Q^{(t)}_0$
 & $B(E2)$ & $Q^{(s)}_0$ & $Q^{(t)}_0$ \\ \hline
   $5/2_1$ \ \ \ \ \ &  
  &     &  35.5 &  99 &       &     &  101 &     \\
 $7/2_1 \rightarrow 5/2_1$ & 383(117)
  & 313 &   8.8 & 131 &  94   & 332 &  142 &  98 \\
 $9/2_1 \rightarrow 5/2_1$ & 153(32)
  &  98 &  -8.5 &  94 &  99   &  97 &   92 & 100 \\
       $\rightarrow 7/2_1$ & 426(149)
  & 287 &       &     &  98   & 283 &      &  98 \\
 $11/2_1 \rightarrow 7/2_1$ & 184(20)
  & 165 & -16.3 &  87 &  99   & 166 &   69 & 101 \\
        $\rightarrow 9/2_1$ & 107(85)
  & 217 &       &     &  97   & 213 &      &  97 \\
 $13/2_1 \rightarrow 9/2_1$ & 62(27)
  & 199 & -25.5 & 102 &  96   & 192 &   98 &  96 \\
        $\rightarrow 11/2_1$ & $4^{+106}_{-4}$
  & 157 &       &     &  94   & 153 &      &  94 \\
 $15/2_1 \rightarrow 11/2_1$ & 102(15)
  & 203 & -22.6 &  77 &  90   & 185 &   47 &  88 \\
        $\rightarrow 13/2_1$ & $<256$
  & 108 &       &     &  88   &  92 &      &  82 \\
 $17/2_1 \rightarrow 13/2_1$ & 70(11)
  & 161 & -15.2 &  46 &  78   &     &   34 &  75 \\
        $\rightarrow 15/2_1$ & 
  &  63 &       &     &  83   &     &      &  72 \\
 $19/2_1 \rightarrow 15/2_1$ & 149(32)
  & 128 &  -6.0 &  17 &  67   & 158 &  9.8 &  76 \\
        $\rightarrow 17/2_1$ & 
  &  51 &       &     &  74   &     &      &  87 \\
 $21/2_1 \rightarrow 17/2_1$ &  0
  & 143 & -18.5 &  50 &  69   &     &   29 &  66 \\
        $\rightarrow 19/2_1$ &
  &  27 &       &     &  59   &     &      &  67 \\
 $23/2_1 \rightarrow 19/2_1$ &
  & 122 &  -8.8 &  23 &  63   &     &   13 &  68 \\
        $\rightarrow 21/2_1$ &
  &  22 &       &     &  58   &     &      &  87 \\
 $25/2_1 \rightarrow 21/2_1$ & 90(8)
  &  21 &  14.4 & -36 &  26   &     &  -39 &  17 \\
        $\rightarrow 23/2_1$ &
  & 0.4 &       &     &  8.6  &     &      &  8.6 \\
 $27/2_1 \rightarrow 23/2_1$ &
  &  56 &   5.1 & -12 & 42    &     &  -5.1 & 52 \\
        $\rightarrow 25/2_1$ &
  &  16 &       &     &  58   &     &       &  50 \\
 $29/2_1 \rightarrow 25/2_1$ & 0+84
  &  46 &   2.2 & -5.2 & 37   &     &  -1.4 & 43 \\
        $\rightarrow 27/2_1$ &
  &  15 &       &     &  60   &     &       &  35 \\
 $31/2_1 \rightarrow 27/2_1$ & $>42$
  &  57 &   2.5 &  -6 &  42   &     &  -4.6 & 42 \\
        $\rightarrow 29/2_1$ &
  & 5.2 &       &     &  37   &     &       &  28 \\
 $1/2_1 \rightarrow 5/2_1$ & $<10$
  & 7.3 &       &     &       & 7.0 &       &     \\
 $3/2_1 \rightarrow 5/2_1$ & 21(6)
  & 3.5 &       &     &       & 4.1 &       &     \\
       $\rightarrow 7/2_1$ & 0.26(12)
  & 0.7 &      &     &       & 4.8 &       &     \\
% $19/2_2 \rightarrow 15/2_1$ &
%  &  81 &  19.1 & -55 &  54   &     &    50 &  37 \\
%        $\rightarrow 17/2_1$ & 
%  & 6.5 &       &     &  27   &     &       &  17 \\
   \hline
\end{tabular}
\end{center}
\end{table}
\end{footnotesize}
%=========================================================

  High-spin states of the $^{49}_{24}$Cr$_{25}$ nucleus were identified
 by recent highly efficient experiments \cite{Cameron2,OLeary}.
 In Fig. \ref{fig11}, we compare calculated energy levels of $^{49}$Cr
 with the observed ones.  Our model well describes the energies of the
 yrast states as a whole.  There are slight deviations
 for the high-spin states $23/2^-_1$, $25/2^-_1$ and $29/2^-_1$.
 Namely, our model does not reproduce enough the observed magnitudes
 of the stagger from $21/2^-_1$ to $23/2^-_1$ and from $25/2^-_1$
 to $27/2^-_1$ in the spin-energy graph. The full $fp$ shell model
 calculation with the KB3 interaction \cite{Martinez2} correctly
 reproduces the inverse order of these pair levels.
 This is another question to our model, in contrast with the success
 in $^{47}$Ti and $^{49}$V.
 There is a confusing difference between the spin-energy graphs of
 the cross conjugate nuclei $^{47}$V and $^{49}$Cr. Their staggering
 gaits up to $19/2^-_1$ are somewhat different between $^{47}$V and
 $^{49}$Cr, {\it i.e.}, the energies of the pair states $(4J+1)/2^-$
 and $(4J+3)/2^-$ are close by in $^{47}$V but are different in
 $^{49}$Cr, which is well traced by our model.
  Above $19/2^-_1$, our model predicts similar staggering gaits
 for $^{47}$V and $^{49}$Cr (which suggests the dominance of the
 configurations $(f_{7/2})^{7p}$ and $(f_{7/2})^{7h}$), while the two
 experimental graphs do not show any regular behavior. This is contrary
 to the fact that our model well reproduces the yrast bands of the
 cross-conjugate odd-odd nuclei $^{46}$V and $^{50}$Mn (see (I)).
 The staggering gaits from $25/2^-_1$ to $27/2^-_1$ and
 from $29/2^-_1$ to $31/2^-_1$ are asymmetrical between $^{47}$V and
 $^{49}$Cr. The subtle but strange behaviors are curious for us to
 understand, though the delicate difference in the interaction matrix
 elements from the realistic effective interaction KB3 could affect
 the staggering in our model. The interesting fine structure should
 be investigated further experimentally and theoretically.

 Electric quadrupole properties of the yrast states in $^{49}$Cr are
 listed in Table \ref{table4}.  The results of our model for $B(E2)$,
 $Q^{(s)}_0$ and $Q^{(t)}_0$ are similar to those of Ref.
 \cite{Martinez2}.  The calculated $B(E2)$ values are roughly
 consistent with the observed ones.  The calculated spectroscopic
 quadrupole moment shows a difference between the two sets of the
 states ($17/2^-_1$, $21/2^-_1$) and ($19/2^-_1$, $23/2^-_1$),
 and indicates a distinct change of the structure above $23/2^-_1$
 in the yrast band.  This feature is different from that of the cross
 conjugate nucleus $^{47}$V but similar to that of $^{49}$V,
 which suggests the contribution of the upper orbits above $f_{7/2}$.
 
%===  Fig. 12 ========================================================
\begin{figure}[b]
\begin{center}
    \epsfig{file=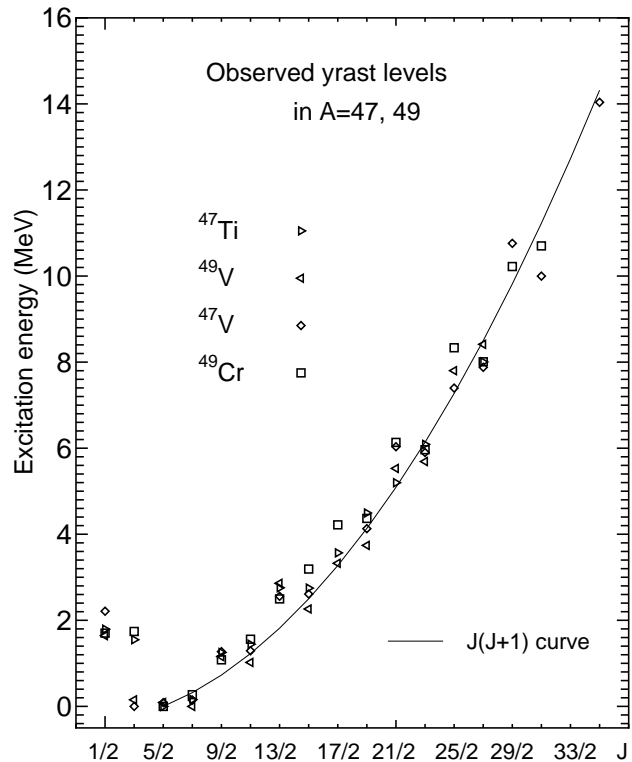,width=0.6\textwidth}
\caption{Experimental spin-energy relation in the yrast bands
         of $A$=47 and $A$=49 nuclei.}
\label{fig12}
\end{center}
\end{figure}
%====================================================================

 In Fig. \ref{fig12}, we summarize the observed yrast levels of
 $^{47}$Ti, $^{47}$V, $^{49}$V and $^{49}$Cr in the spin-energy graph.
 The yrast levels appear to be scattered around the $J(J+1)$ curve of
 the rotation model, where the static moment of inertia is fixed as
 $I_{mom}$=11 MeV$^{-1}$ and the energy of the $5/2^-_1$ state is set
 to be zero.  This figure indicates a similarity at the beginning of
 the band on the $5/2^-_1$ state between $^{47}$V and $^{49}$Cr and
 also between $^{47}$V and $^{47}$Ti. 
 Mart\'{\i}nez-Pinedo {\it et al.} \cite{Martinez2} analyzed old data
 of $^{47}$V and $^{49}$Cr by different curves of the $J(J+1)$ law.
 In our model, the spin-energy graphs of $^{47}$V and $^{49}$Cr do not
 display a very large difference, though the spectroscopic quadrupole
 moment shows differences as mentioned above ($Q_{spec}$ shows
 a drastic change at $25/2^-_1$ in $^{49}$Cr but not in $^{47}$V).
 The asymmetrical positions of ($25/2^-_1$, $27/2^-_1$) and
 ($29/2^-_1$, $31/2^-_1$) in $^{47}$V and $^{49}$Cr mentioned above
 are visible in Fig. \ref{fig12}.

\section{Results in $^{51}$Mn and $^{51}$Cr}

   The extended $P+QQ$ model is expected to be applicable to $A=51$
 nuclei from Fig. \ref{fig3}.  In this section, we show calculated
 results in $^{51}$Mn and $^{51}$Cr.
   
%===  Fig. 13 ========================================================
\begin{figure}[b]
\begin{center}
    \epsfig{file=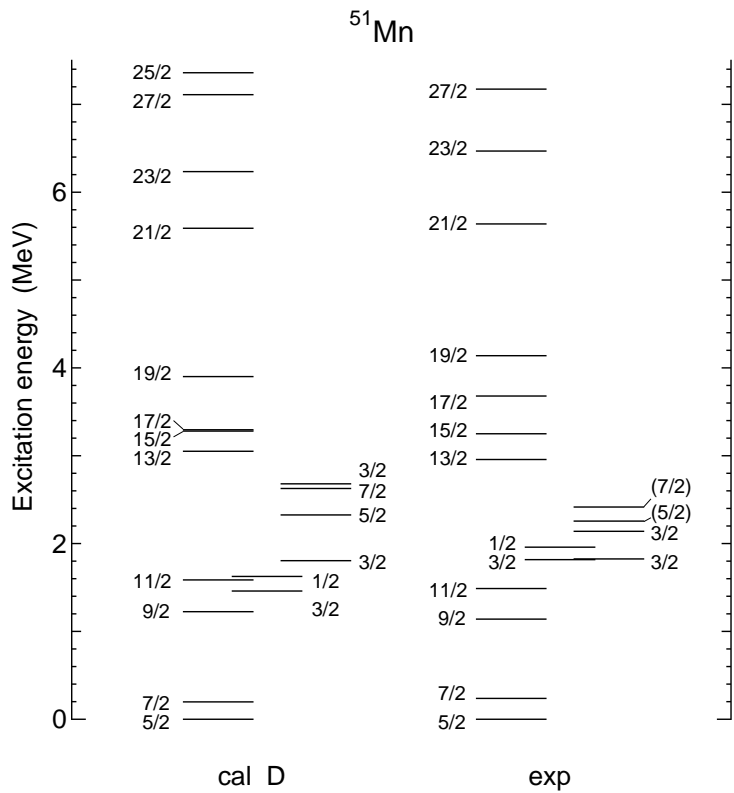,width=0.7\textwidth}
\caption{Comparison of energy levels between theory and experiment
        in $^{51}$Mn.}
\label{fig13}
\end{center}
\end{figure}
%====================================================================

   Figure \ref{fig13} illustrates the comparison of energy levels
 between the modified model (D) and experiment in $^{51}$Mn.
 Our model reproduces the collective yrast levels on $5/2^-_1$ at good
 energy and in the correct order of spins, except that the space
 between $15/2^-_1$ and $17/2^-_1$ is too narrow.  (The awkward
 reproduction of energy in the middle spin region in our calculation
 is similar to that in $^{48}$Cr.)  The additional monopole
 terms improve the positions of the low-spin levels $3/2^-_1$ and
 $1/2^-_1$, the spins of which are smaller than $J=5/2$ of
 the band-head state.  The excitation energies of observed non-yrast
 levels are nearly reproduced.
 
   Let us list calculated electric quadrupole properties of the yrast
 states in Table \ref{table5}, in order to see the structure of
 $^{51}$Mn. Table \ref{table5} shows that the yrast states from $5/2^-$
 to $15/2^-$ are connected by large values of $B(E2:J \rightarrow J-2)$
 and $B(E2:J \rightarrow J-1)$.  We can regard these states as members
 of a very collective band.  The spectroscopic quadrupole moment
 $Q_{spec}$ abruptly changes the sign at $17/2^-_1$, which suggests a
 structure change similar to that at $10^+_1$ in $^{50}$Cr (see (I)). 
 The high-spin yrast states on $17/2^-_1$ that have still relatively
 large $B(E2)$ values can be regarded to belong to a different-natured
 band. The awkward gait from $15/2^-_1$ to $17/2^-_1$ in the calculated
 result reflects the abrupt change in the structure.  We can see
 such a sign in the observed level scheme of $^{51}$Mn.
 
%===================  Mn(51)  ====================
\begin{footnotesize}
\begin{table}[b]
\caption{Calculated electric quadrupole properties of the yrast states
 in $^{51}$Mn: ${B(E2)}$ in $e^2$ fm$^4$; $Q_{spec}$ in $e$ fm$^2$.}
\label{table5}
\begin{center}
\begin{tabular}{rccccc} \hline
  $J_1$ & $Q_{spec}$ & $\rightarrow J^\prime_1$  & $B(E2)$
   &  $\rightarrow J^\prime_1$ &  $B(E2)$  \\ \hline
 $1/2_1$ & 
          & $\rightarrow  5/2_1$ & 39  & $\rightarrow  3/2_1$ & 454 \\
 $3/2_1$ & -22.0
          & $\rightarrow  7/2_1$ &  7  & $\rightarrow  5/2_1$ &  15 \\
 $5/2_1$ & 35.1
          &                      &     &                      &     \\
 $7/2_1$ &  8.0
          &                      &     & $\rightarrow  5/2_1$ & 308 \\
 $9/2_1$ & -7.7
          & $\rightarrow  5/2_1$ &  88 & $\rightarrow  7/2_1$ & 239 \\
 $11/2_1$ & -12.2
          & $\rightarrow  7/2_1$ & 148 & $\rightarrow  9/2_1$ & 194 \\
 $13/2_1$ & -20.6
          & $\rightarrow  9/2_1$ & 167 & $\rightarrow 11/2_1$ & 116 \\
 $15/2_1$ & -8.6
          & $\rightarrow 11/2_1$ & 150 & $\rightarrow 13/2_1$ &  84 \\
 $17/2_1$ & 61.1
          & $\rightarrow 13/2_1$ & 1.0 & $\rightarrow 15/2_1$ & 0.0 \\
 $19/2_1$ & 35.3
          & $\rightarrow 15/2_1$ & 7.8 & $\rightarrow 17/2_1$ & 116 \\
 $21/2_1$ & 25.5
          & $\rightarrow 17/2_1$ &  25 & $\rightarrow 19/2_1$ & 146 \\
 $23/2_1$ & 11.7
          & $\rightarrow 19/2_1$ &  49 & $\rightarrow 21/2_1$ &  91 \\
 $25/2_1$ & 23.7
          & $\rightarrow 21/2_1$ &  24 & $\rightarrow 23/2_1$ &  14 \\
 $27/2_1$ & 37.2
          & $\rightarrow 23/2_1$ &  32 & $\rightarrow 25/2_1$ & 0.0 \\
   \hline
\end{tabular}
\end{center}
\end{table}
\end{footnotesize}
%=========================================================

%===  Fig. 14 ========================================================
\begin{figure}
\begin{center}
    \epsfig{file=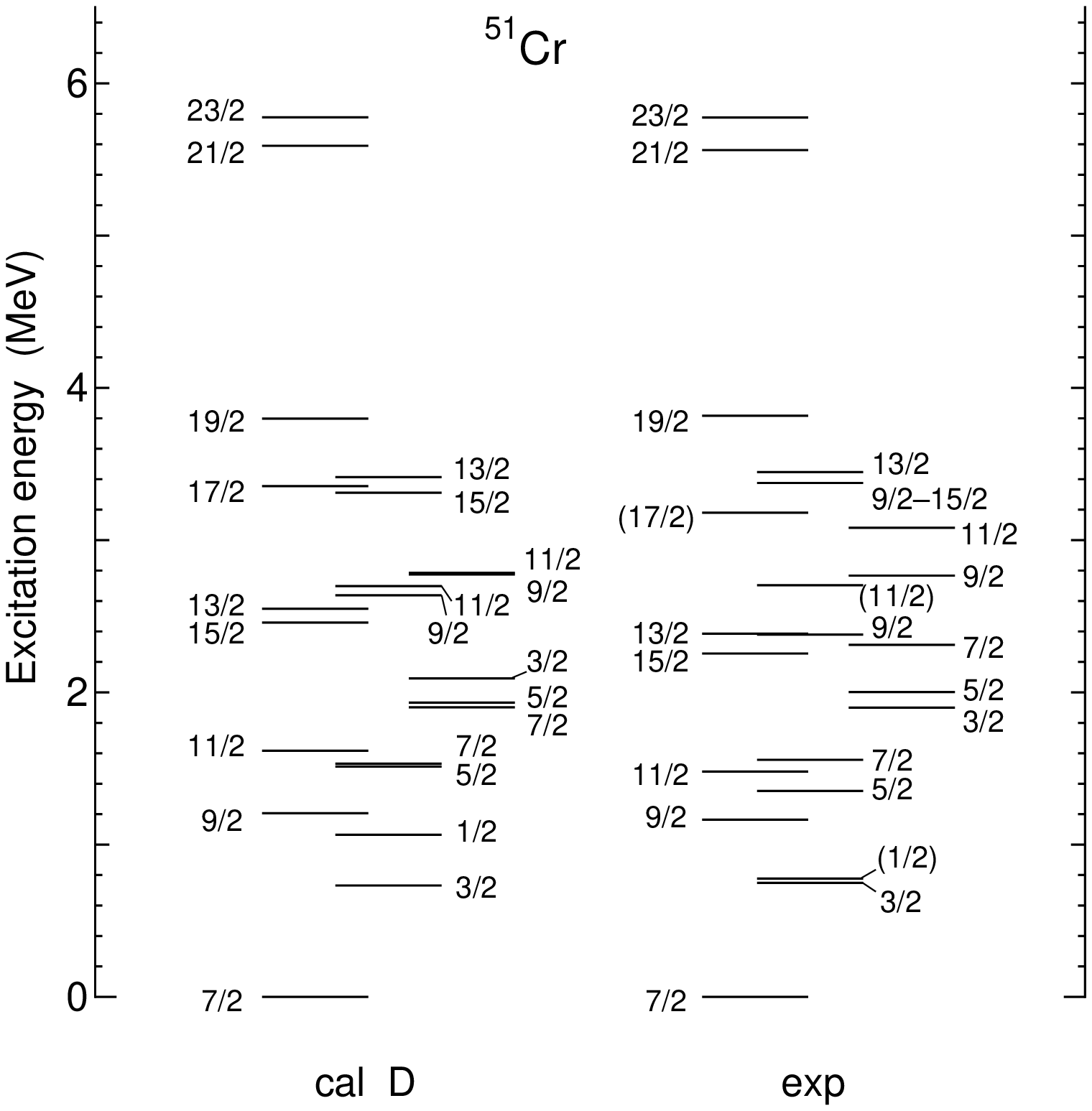,width=0.7\textwidth}
\caption{Comparison of energy levels between theory and experiment
        in $^{51}$Cr.}
\label{fig14}
\end{center}
\end{figure}
%====================================================================

%===================  Cr(51)  ====================
\begin{footnotesize}
\begin{table}
\caption{Calculated electric quadrupole properties of the yrast states
 in $^{51}$Cr: ${B(E2)}$ in $e^2$ fm$^4$; $Q_{spec}$ in $e$ fm$^2$.}
\label{table6}
\begin{center}
\begin{tabular}{rccccc} \hline
  $J_1$ & $Q_{spec}$ & $\rightarrow J^\prime_1$  & $B(E2)$
   &  $\rightarrow J^\prime_1$ &  $B(E2)$  \\ \hline
 $1/2_1$ & 
          &                      & 39  & $\rightarrow  3/2_1$ & 424 \\
 $3/2_1$ & -20.4
          & $\rightarrow  7/2_1$ & 0.0 &                      &     \\
 $5/2_1$ &  24.5
          & $\rightarrow  1/2_1$ & 0.0 & $\rightarrow  3/2_1$ & 0.0 \\
 $5/2_1$ & 
          & $\rightarrow  9/2_1$ & 3.9 & $\rightarrow  7/2_1$ &  56 \\
 $7/2_1$ &  35.2
          &                      &     &                      &     \\
 $9/2_1$ &  15.1
          &                      &     & $\rightarrow  7/2_1$ & 261 \\
 $11/2_1$ &  1.8
          & $\rightarrow  7/2_1$ &  69 & $\rightarrow  9/2_1$ & 227 \\
 $13/2_1$ & -7.6
          & $\rightarrow  9/2_1$ &  93 & $\rightarrow 11/2_1$ & 176 \\
 $15/2_1$ & 33.2
          & $\rightarrow 11/2_1$ &  48 & $\rightarrow 13/2_1$ &  22 \\
 $17/2_1$ & 17.4
          & $\rightarrow 13/2_1$ &  62 & $\rightarrow 15/2_1$ &  83 \\
 $19/2_1$ & 12.0
          & $\rightarrow 15/2_1$ &  28 & $\rightarrow 17/2_1$ &  57 \\
 $21/2_1$ & 0.8
          & $\rightarrow 17/2_1$ &  54 & $\rightarrow 19/2_1$ &  36 \\
 $23/2_1$ & 7.8
          & $\rightarrow 19/2_1$ &  61 & $\rightarrow 21/2_1$ &  18 \\
   \hline
\end{tabular}
\end{center}
\end{table}
\end{footnotesize}
%=========================================================

  Figure \ref{fig14} shows the results in $^{51}$Cr. The energy levels
 observed in $^{51}$Cr are excellently described by our model.
 The extended $P+QQ$ model with the additional monopole terms
 reproduces not only the ground-state band but also the apparently
 second band, at good energy and in the correct order of spins.
 The inverse orders of ($13/2^-_1$, $15/2^-_1$) and
 ($1/2^-_1$, $3/2^-_1$) in $^{51}$Cr are reproduced by our model.
 The calculated energies of the $J^-_3$ states roughly correspond to
 those of the observed third band on $3/2^-_2$. 

  Calculated electric quadrupole properties of the yrast states of
 $^{51}$Cr are shown in Table \ref{table6}.  This table indicates
 that there are a very collective band from $7/2^-_1$ to $13/2^-_1$
 and a different-natured band on $15/2^-_1$.  We have a similar
 structure change at middle spin both in $^{51}$Mn and $^{51}$Cr,
 though the turning point is different and the behavior of the
 spectroscopic quadrupole moment $Q_{spec}$ is different in the two
 nuclei.  This may be the same kind of band crossing as that
 in $^{50}$Cr or $^{48}$Cr.  The backbending is common to even-$A$
 and odd-$A$ nuclei in the middle of $f_{7/2}$ shell.

  Although we omit the results in $^{51}$V and $^{51}$Ti, the extended
 $P+QQ$ model with the additional monopole terms well describes these
 nuclei.

%====================================================================
\section{Concluding remarks}

   We have shown that the $P_0+P_2+QQ+V^0_{\pi \nu}$ interaction
 proposed in Refs. \cite{Hasegawa,Kaneko,Hasegawa2} is very much
 improved, if the average monopole field $V^0_{\pi \nu}$ is
 modified by adding small monopole terms.  The modified model well
 reproduces the experimental binding energies, energy levels and
 $B(E2)$ between the yrast states not only in even-$A$ nuclei
 but also odd-$A$ nuclei with $A$=44-51.  The concrete results support
 the previous discussions developed by Zuker and co-workers:
  (1) the monopole components of the $T=0$ and $T=1$ interaction
 matrix elements play important roles in nuclear spectroscopy
 \cite{Caurier,Caurier2,Martinez,Martinez2,Poves2};
  (2) the residual part of a realistic effective interaction obtained
 after extracting the monopole terms is dominated by the $P_0$ and $QQ$
 forces  \cite{Dufour}.
   Our functional interaction has clarified the different roles of
 the average monopole field $V^0_{\pi \nu}$ (which contributes to
 the large binding energy and to the energy difference between
 the $T=0$ and $T=1$ states in $N = Z$ nuclei
 \cite{Hasegawa,Kaneko,Hasegawa2,Kaneko2}) and the additional monopole
 terms (which affect spectroscopy).  The calculations recommend
 including the $P_2$ force in quantitative description.
 
   The present modification of the $P_0+P_2+QQ+V^0_{\pi \nu}$
 interaction leaves still a tiny room for improvement.  For instance,
 we have seen the somewhat awkward reproduction of the middle-spin
 levels in the band crossing region in some nuclei such as $^{51}$Mn.
  We have considered only the simplified forms of the $T=0$ and $T=1$
 diagonal interactions and neglect $J$-dependence of them,
 in this paper.  On the other hand, some $J$-dependent miner changes
 in the interaction matrix elements
 $\langle f_{7/2}f_{7/2}JT | V | f_{7/2}f_{7/2}JT \rangle$
 are added to the monopole terms in the KB3 interaction.  If we adopt
 such miner corrections for $\Delta k^\tau(f_{7/2}f_{7/2}J)$ in Eqs.
 (\ref{eq:4}) and (\ref{eq:8}), we shall possibly get better results.
  We made the present calculations without changing the scaling of
 the force strengths, since the results do not demand the change.
 We have simply neglected the $A$-dependence of the additional
 monopole terms.
 
   The calculations in this paper (and (I)) demonstrate the quality of
 the $P_0+P_2+QQ+V^0_{\pi \nu}$ interaction modified by the
 additional monopole terms (or without the modification in some cases)
 to be good enough for practical use to discuss the nuclear structure.
 Our interaction is flexible in changing the model space (for instance
 possible extension to the two major-shell space) and is
 applicable to regions where any reliable effective interaction is not
 available.  Our functional effective interaction is not determined
 in terms of individual interaction matrix elements but has only a few
 parameters (8 parameters and single-particle energies in this paper).
 It is interesting to apply the functional effective interaction to
 various nuclei including those in the proton-drip line.

\begin{ack}
 The authors are very grateful to Dr. T. Mizusaki for taking interest
in our work and giving us a hint to improve our computer code.
\end{ack}

%================================================================
%\newpage

%*************************************************
\end{document}